\documentclass[12pt,amstex,aps,amsfonts,prsty]{article}
\usepackage[dvips]{graphicx}
\usepackage{amssymb}
\usepackage{amsmath}
\usepackage{a4wide}
\usepackage{citesort}
\renewcommand{\baselinestretch}{1.1}
\begin{document}
\begin{center}
{\bf Intrinsic Friction of Monolayers Adsorbed on Solid Surfaces.}
\end{center}
{\bf O.B{\'e}nichou$^{1}$, A.M.Cazabat$^1$,  J.De Coninck$^{2}$,
 M.Moreau$^3$ and  G.Oshanin$^{3,4}$\\}
{$^1$  Laboratoire de Physique de la Mati{\`e}re Condens{\'e}e, \\
Coll{\`e}ge de France, 11 Place M.Berthelot, 75252 Paris Cedex 05, France\\
}
{$^2$ Centre de Recherche en Mod{\'e}lisation Mol{\'e}culaire, \\
Universit{\'e} de Mons-Hainaut, 20 Place du Parc, 7000 Mons, Belgium\\
}
{$^3$  Laboratoire de Physique Th{\'e}orique des Liquides, \\
Universit{\'e} Paris 6, 4 Place Jussieu, 75252 Paris, France\\
}
{$^4$  Max-Planck-Institut f\"ur Metallforshung\\
Heisenbergstr. 3, 70569 Stuttgart, Germany\\
}

{\bf ABSTRACT}

\vspace{0.2in}

We overview recent results on
 intrinsic frictional properties
 of adsorbed monolayers,
composed of mobile hard-core particles
undergoing continuous exchanges with a vapor phase.
In terms of a dynamical master equation approach
we determine the velocity of
a biased impure molecule -
the tracer particle (TP),
constrained to move inside the adsorbed monolayer probing its
frictional properties,
define the frictional forces exerted by the monolayer on the TP,
as well as the
particles density distribution in the monolayer.

\vspace{0.3in}

{\bf INTRODUCTION}

\vspace{0.2in}

Monolayers emerging on solid surfaces  exposed to
a vapor phase are important in different backgrounds, including
such
technological and material processing operations as, e.g.,
coating, gluing or lubrication. Knowledge of their intrinsic frictional properties is
important for  understanding of different transport processes taking place
within molecular films, film's stability,
as well as spreading of ultrathin
liquid films on solid surfaces \cite{spreading1},
spontaneous or forced dewetting of monolayers
\cite{aussere,dewetting}
or island formation  \cite{islands}.

Since the early works of Langmuir,
much effort has been invested in the analysis of the
equilibrium properties of
the adsorbed films
\cite{surf}.
Significant analytical results have been obtained
predicting  different phase transitions and ordering
phenomena.
As well,
some approximate results have been obtained for
both dynamics of isolated non-interacting adatoms
on corrugated surfaces and collective
diffusion, describing spreading of the macroscopic
density fluctuations in
 interacting adsorbates
\cite{wahn}.

Another important aspect of dynamical behavior concerns tracer
 diffusion in adsorbates,
which is observed experimentally in
STM or field ion measurements
 and provides
a useful information about adsorbate's viscosity or intrinsic
friction. This problem is not only a challenging question
in its own right due to emerging non-trivial, essentially cooperative behavior,
but is also
crucial for understanding of various dynamical processes taking
place on solid surfaces. Most of available theoretical
studies of tracer diffusion in adsorbed layers
(see, e.g.,
\cite{kehr})
exclude, however,
the
 possibility of
particles exchanges with the vapor, which limits their application to the analysis of behavior in realistic systems.

Here we focus on this important problem and provide a theoretical
description of the properties of tracer diffusion in adsorbed
monolayers in contact with a vapor phase. More specifically, the
system we consider consists of (a) a solid substrate, which is
modelled in a usual fashion as a regular
 lattice of adsorption sites; (b)
a monolayer of adsorbed, mobile
hard-core particles in contact with a vapor
and (c) a single hard-core tracer
particle (TP). We suppose that the monolayer particles move randomly
along the lattice
by performing symmetric
hopping motion between
the neighboring lattice sites, which process
 is constrained by mutual hard-core interactions, and
 may desorb from and adsorb onto the lattice from the vapor
with  some prescribed rates dependent on the vapor
pressure, temperature
 and the interactions with the solid substrate.
In contrast, the
tracer particle is constrained to move
along the
lattice only,
(i.e. it can not desorb
to the vapor), and  is
subject to
a constant external force of an arbitrary magnitude $E$. Hence,
the TP performs
a biased random walk,
constrained by the  hard-core
interactions with the
monolayer particles,
and always remains within
the monolayer, probing
its frictional properties.

The questions  we address here are the following: First, we aim to
determine the  force-velocity relation, i.e., the dependence of
the TP terminal velocity $V_{tr}(E)$ on the magnitude of the
applied external force. Next, we study the form of the
force-velocity  relation in the limit of a vanishingly small
external bias. This allows us, in particular, to show that the
frictional force exerted on the TP by the monolayer particles is
viscous, and to evaluate the corresponding friction coefficient.
Lastly, we analyze how the biased TP perturbs the particles
density distribution in the monolayer;  we proceed to show that
there are stationary density profiles around the TP, which mirror
a remarkable cooperative behavior. We also consider the case of
monolayers sandwiched between two solid surfaces and demonstrate
that here the cooperative phenomena are dramatically enhanced.
Detailed account of these results can be found in the original
papers   \cite{benichou,physa,prl,benichouPRB}.

We finally remark that our analysis can be viewed from a different perspective.
On the one hand, the system under study
represents  a certain generalization of the  "tracer diffusion in
a
hard-core lattice gas" problem (see, e.g., \cite{kehr}) to the case where the random walk
performed by the TP is {\it biased} and the number of particles in
the monolayer is {\it not explicitly conserved}, due to exchanges with
the reservoir. We recall that even this, by now classic model
constitutes a many-body problem for which no exact general expression of
the tracer diffusion coefficient $D_{tr}$ is known.
On the other hand, our model provides a novel example
 of the so called ``dynamical percolation'' models (see, e.g. Ref.\cite{nouspercol} for a review),
in which the monolayer particles act  as a fluctuating
environment, which hinders the motion of an impure molecule, say,
a charge carrier.
  Lastly, we note that the model under study can be thought of as
some simplified picture of the
stagnant layers emerging in liquids being in contact with a solid body. It is well
known (see, e.g. Ref.\cite{lyklema}) that liquids in close vicinity
 of a solid interfaces - at distances within a few molecular diameters, do possess completely
different physical properties compared to those characterizing the bulk phase.
In this "stagnant" region, in which an intrinsically disordered liquid phase is spanned by and contends
with the ordering potential of the solid,
liquid's viscosity is drastically enhanced
and transport processes  are
essentially hindered. Thus our model can be viewed
as a two-level approximate model of this challenging physical system,
in which
the reservoir mimics the bulk fluid phase with very rapid transport,
while the adsorbed monolayer represents the stagnant layer emerging on the solid-liquid interface.

 \vspace{0.3in}

{\bf THE MODEL AND THE EVOLUTION EQUATIONS}

\vspace{0.2in}

Consider a
two-dimensional
solid surface with some concentration of adsorption sites,
which is brought
in
contact with a reservoir containing identic,
electrically neutral particles  - a vapor phase (Fig.\ref{reseau}),
maintained at a constant pressure. For simplicity of exposition, we assume here
that adsorption sites form a regular square lattice
of spacing $\sigma$.
We suppose next that the reservoir particles
may adsorb
onto any vacant adsorption site at a fixed rate $f/\tau^*$, which rate
depends on the vapor
pressure and the energy gain due to the adsorption event. Further on,
the adsorbed particles may move randomly along the lattice by
hopping at a rate $1/ 4 \tau^*$ to any of $4$
neighboring adsorption sites,
which process is
constrained by hard-core exclusion preventing multiple occupancy of any of the sites.
Lastly, the adsorbed particles may
 desorb from the lattice  back to the reservoir
at rate $g/\tau^*$, which is dependent on the barrier against desorption.
Both $f$ and $g$ are site and environment independent.

\begin{figure}[ht]
\begin{center}
\includegraphics*[scale=0.4]{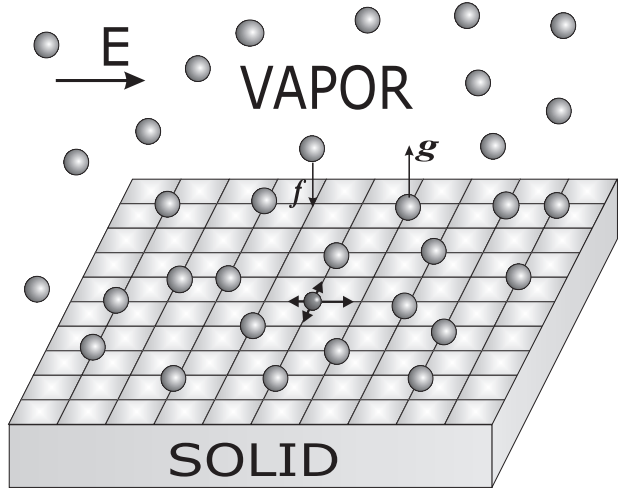}
\end{center}
\caption{\label{reseau} {\small Adsorbed monolayer in contact with a vapor.
Grey spheres denote the
monolayer (vapor) particles; the smaller black sphere
stands for the driven TP.}}
\end{figure}

Further on, at $t = 0$ we introduce at the lattice origin an extra
hard-core particle, whose motion we would like to follow; position
of this particle at time $t$ is denoted as ${\bf R}_{tr}$.  We
stipulate that the TP  can not desorb from the lattice and that
it is subject to some external driving force, which favors its
jumps into a preferential direction. Physically, this situation may be realized 
if the TP is charged, while the monolayer particles are neutral, and the whole system
is subject to external electric field. 
The TP transition
probabilities are determined by:
\begin{equation}
p_\nu=\frac{\exp\Big[\frac{\beta}{2}({\bf E \cdot e}_{\boldsymbol
\nu})\Big]}{\sum_{\mu}\exp\Big[\frac{\beta}{2}({\bf E \cdot e}_{\boldsymbol \mu})\Big]},
\label{defp}
\end{equation}
where $\beta$ is the reciprocal temperature, $({\bf E \cdot e})$ stands for the scalar product,
the charge of the TP is set equal to unity and
the sum with the subscript $\mu$ denotes summation over all possible
orientations
of the vector ${\boldsymbol e_\mu}$; that is, $\mu = \{\pm1,\pm2\}$.
  We suppose here that the external force ${\bf E}$
 is oriented according to the unit vector
${\bf e_1}$.

In the general $d$-dimensional case, the time evolution of $P({\bf R_{tr}},\eta;t)$ -
the joint probability of finding
at time $t$ the TP at the site ${\bf R_{tr}}$
and all adsorbed particles in the configuration $\eta$, where $\eta\equiv
\{\eta({\bf R})\}$ denotes the entire set of the occupation variables of different sites,
obeys the following Master equation
\cite{benichou,physa,prl,benichouPRB}
\begin{eqnarray}
&&\partial_tP({\bf R_{tr}},\eta;t)=
\frac{1}{2d\tau^*}\sum_{\mu=1}^d\;\sum_{{\bf r}\neq{\bf R_{tr}}-{\bf e}_{\boldsymbol \mu},{\bf R_{tr}}}  \;
\Big\{ P({\bf R_{tr}},\eta^{{\bf r},\mu};t)-P({\bf R_{tr}},\eta;t)\Big\}\nonumber\\
&+&\frac{1}{\tau}\sum_{\mu}p_\mu\Big\{\left(1-\eta({\bf R_{tr}})\right)P({\bf R_{tr}}-{\bf e}_{\boldsymbol \mu},\eta;t)
-\left(1-\eta({\bf R_{tr}}+{\bf e}_{\boldsymbol \mu})\right)P({\bf R_{tr}},\eta;t)\Big\}\nonumber\\
&+&\frac{g}{\tau^*}\sum_{{\bf r}\neq {\bf R_{tr}}} \;\Big\{\left(1-\eta({\bf r})\right)P({\bf R_{tr}},
\hat{\eta}^{{\bf r}};t)-\eta({\bf r})P({\bf R_{tr}},\eta;t)\Big\}\nonumber\\
&+&\frac{f}{\tau^*}\sum_{{\bf r}\neq{\bf R_{tr}}} \;\Big\{\eta({\bf r})P({\bf R_{tr}},\hat{\eta}^{{\bf r}};t)
-\left(1-\eta({\bf r})\right)P({\bf R_{tr}},\eta;t)\Big\}.
\label{eqmaitresse}
\end{eqnarray}
where $\eta^{{\bf r},\nu}$ is the configuration obtained from
$\eta$ by the Kawasaki-type exchange of the occupation variables
of two neighboring sites ${\bf r}$ and ${\bf r+e}_{\boldsymbol
\nu}$, and ${\hat \eta}^{\bf r}$ - a configuration obtained from
the original $\eta$ by the replacement $\eta({\bf r}) \to
1-\eta({\bf r})$, which corresponds to the Glauber-type flip of
the occupation variable due to the adsorption/desorption events.

The mean
velocity $V_{tr}(t)$
of the TP can be obtained by multiplying both sides of Eq.(\ref{eqmaitresse}) by
$({\bf R_{tr} \cdot e_1})$ and summing over all possible configurations $({\bf R_{tr}},\eta)$, which yields
\begin{equation}
V_{tr}(t)\equiv\frac{d}{dt} \; \sum_{{\bf R_{tr}},\eta} ({\bf R_{tr} \cdot e_1})P({\bf R_{tr}},\eta;t) =
\frac{\sigma}{\tau}\Big\{p_1  \Big(1-k({\bf e_1};t)\Big)-p_{-1} \Big(1-k({\bf e_{-1}};t)\Big)\Big\},
\label{vitesse}
\end{equation}
where
\begin{equation}
k({\boldsymbol \lambda};t)\equiv\sum_{{\bf R_{tr}},\eta}\eta({\bf R_{tr}}+{\boldsymbol \lambda})P({\bf R_{tr}},\eta;t)
\label{defk}
\end{equation}
is the probability of having at time t an
adsorbed particle
at position ${\boldsymbol \lambda}$,
defined in the frame of reference moving with the TP.
In other words,  $k({\boldsymbol \lambda};t)$
 can be thought of as being
the density
profile in the adsorbed monolayer as seen from the moving TP.

Note that Eq.(\ref{vitesse}) signifies that
the TP velocity is
dependent on the monolayer particles density in its
immediate vicinity.
If the monolayer is perfectly stirred, i.e., if
$k({\boldsymbol \lambda};t) = \rho_s$ everywhere, where $\rho_s = f/(g+f)$ is the Langmuir adsorption isotherm \cite{surf},
  one would obtain
from Eq.(\ref{vitesse}) a trivial mean-field result
$V_{tr}^{(0)}=(p_1-p_{-1})(1-\rho_s) (\sigma/\tau)$, which states
that the only effect of the medium on the TP dynamics is that its
jump time $\tau$ is renormalized by $(1-\rho_s)^{-1}$. We proceed
to show, however, that $k({\boldsymbol \lambda};t)$ is different
from  $\rho_s$ everywhere, except for
 $|\boldsymbol \lambda|\to\infty$. This means that the TP strongly
perturbs the monolayer.

The method of solution of Eq.(\ref{eqmaitresse}) and
(\ref{vitesse}) has been amply discussed
in \cite{benichou,physa,prl,benichouPRB}. Here we merely present the results.

\vspace{0.3in}

{\bf RESULTS FOR ONE-DIMENSIONAL MONOLAYERS}

\vspace{0.2in}

We focus here
on the one-dimensional lattice, which is
 appropriate to adsorption on polymer chains \cite{10}, and on the limit $t \to \infty$. In this case,
the stationary particles density profile, as seen from steadily moving TP, has
the following form:
\begin{equation}
\label{dprofiles}
k_{n} \equiv k(\lambda) =  \rho_s + K_{\pm} exp\Big(-\sigma |n|/\lambda_{\pm}\Big),
\;\;\; \lambda = \sigma n, \;\;\; n \in Z,
\end{equation}
where the characteristic
lengths $\lambda_{\pm}$ obey
\begin{equation}
\label{lambdas}
\lambda_{\pm}= \; \mp \; \sigma \; ln^{-1}\Big[
\frac{A_{1} + A_{-1} + 2 (f + g) \mp \sqrt{\Big(A_{1} + A_{-1} + 2 (f + g)\Big)^2 - 4 A_{1} A_{-1}}}{2 A_{1}}
\Big],
\end{equation}
the amplitudes $K_{\pm}$ are given by
\begin{equation}
K_{+} = \rho_s \frac{A_{1} - A_{-1}}{A_{-1} - A_{1} \exp(- \sigma/\lambda_{+})}, \;\;\; K_{-} = \rho_s \frac{A_{1} -
A_{-1}}{A_{-1} \exp(- \sigma/\lambda_{-}) - A_{1} },
\end{equation}
the terminal velocity $V_{tr} = \sigma (A_{1} - A_{-1})/2 \tau^*$,
while $A_{1}$ and $A_{-1}$ obey:
\begin{equation}
\label{o}
A_{1} = 1 + \frac{p_1 \tau^*}{\tau} \Big[1 - \rho_s - \rho_s \frac{A_{1} - A_{-1}}{A_{-1} exp(\sigma/\lambda_{+}) - A_{1}}
\Big],
\end{equation}
and
\begin{equation}
\label{b}
A_{-1} = 1 + \frac{p_{-1} \tau^*}{\tau} \Big[1 - \rho_s - \rho_s \frac{A_{1} - A_{-1}}{A_{-1}  - A_{1} exp(\sigma/\lambda_{-})}
\Big].
\end{equation}

\begin{figure}[ht]
\begin{center}
\includegraphics*[scale=0.5]{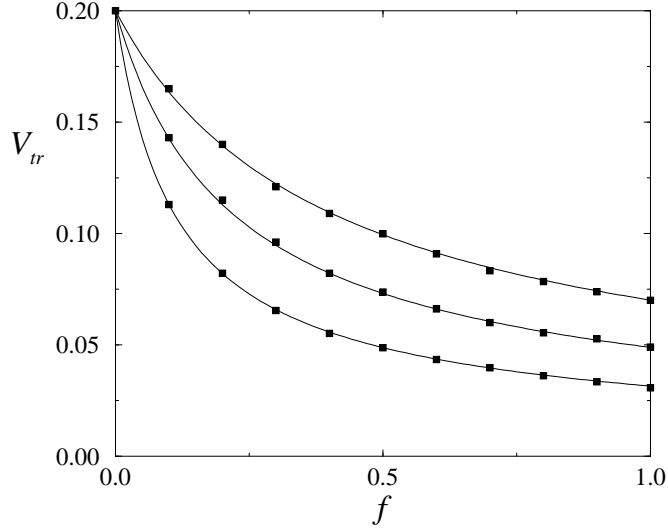}
\end{center}
\caption{\label{vitfig} {\small The TP velocity as a function of
the adsorption probability $f$ at different values of $g$. The TP
hopping probabilities are $p_1 = 0.6$ and $p_{-1} = 0.4$. The
solid lines  give the analytical solution while the squares denote
the Monte-Carlo simulations results. Upper curves correspond to $g
= 0.8$, the intermediate - to $g = 0.5$ and the lower - to $g =
0.3$, respectively.}}
\end{figure}

Note that $\lambda_{-} > \lambda_{+}$, and consequently,  the
local density past the TP approaches its non-perturbed value
$\rho_s$ slower than in front of it.  Next,  $K_{+}$ is always
 positive, while $K_{-} < 0$; this means that
the density profile is a non-monotonous function of $\lambda$ and
is characterized by a jammed region in front of the TP, in which
the local density is higher than $\rho_s$, and a depleted region
past the TP in which the density is lower than $\rho_s$.

\begin{figure}[ht]
\begin{center}
\includegraphics*[scale=0.5]{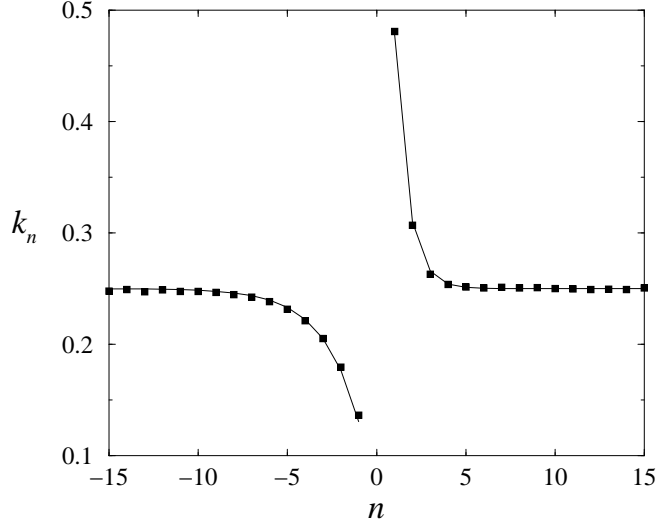}
\end{center}
\caption{\label{proffig} {\small Density profile around stationary
moving TP for $f = 0.1$, $g = 0.3$ and $p = 0.98$. The solid line
is the plot of the analytical solution. Squares are the MC
results.} }
\end{figure}

 For arbitrary values of $p$, $f$ and $g$ the parameters $A_{\pm
1}$, Eqs.(\ref{o}) and (\ref{b}), and consequently, $V_{tr}$ can
be determined only numerically (see Figs.2 to 4). However,
$V_{tr}$ can be found analytically in the limit $E \to 0$. In the
leading in $E$ order, we find
\begin{equation}
\label{st}
V_{tr} \sim \zeta^{-1} E,
\end{equation}
which relation is an analog of the Stokes formula for driven
motion in a 1D adsorbed monolayer undergoing continuous particles
exchanges with the vapor phase and signifies that the frictional
force exerted on the TP by the monolayer particles is $viscous$.
The friction coefficient $\zeta$ is given explicitly by
\begin{equation}
\label{dfriction}
\zeta = \frac{2 \tau}{\beta \sigma^2 (1 - \rho_s)} \Big[1 + \frac{\rho_s \tau^*}{\tau (f + g)}
\frac{2}{1 + \sqrt{1 + 2 (1 + \tau^* (1 - \rho_s)/\tau)/(f + g)}} \Big]
\end{equation}
Note that the friction coefficient in Eq.(\ref{dfriction}) can be
written down as the sum of two contributions $
\zeta=\zeta_{cm}+\zeta_{coop}$. The first one, $\zeta_{cm}=2\tau/
\beta \sigma^2(1-\rho_s)$ is a typical mean-field result and
corresponds to a perfectly homogeneous monolayer. The second one,
\begin{equation}
\zeta_{coop}=\frac{8\tau^*\rho_s}{\beta\sigma^2(1-\rho_s)(f+g)}\frac{1}{1+\sqrt{1 + 2 (1 + \tau^* (1 - \rho_s)/\tau)/(f + g)}},
\end{equation}
has a more complicated origin and reflects a cooperative behavior
emerging in the monolayer, associated with the formation of
inhomogeneous density profiles (see Fig.\ref{proffig}) - the
formation of a ``traffic jam'' in front of the TP and a
``depleted'' region  past the TP (for more details, see
\cite{benichou}).  The characteristic lengths of these two regions
as well as the amplitudes $K_{\pm}$ depend on the magnitude of the
TP velocity; on the other hand, the TP velocity is itself
dependent on the density profiles, in virtue of
Eq.(\ref{vitesse}). This results in an intricate interplay between
the jamming effect of the TP and smoothening of the created
inhomogeneities by diffusive processes. Note also that cooperative
behavior becomes most prominent in  the conserved particle number
limit. Setting $f,g \to 0$, while keeping their ratio fixed (which
insures that $\rho_s$ stays constant), one notices that
$\zeta_{coop}$ gets infinitely large. As a matter of fact, in such
a situation no stationary density profiles around the TP exist;
the size of both the "traffic jam" and depleted regions grow in
proportion to the TP mean displacement $\overline{X_{tr}(t)} \sim
\sqrt{t}$ \cite{bur2}.

\begin{figure}[ht]
\begin{center}
\includegraphics*[scale=0.5]{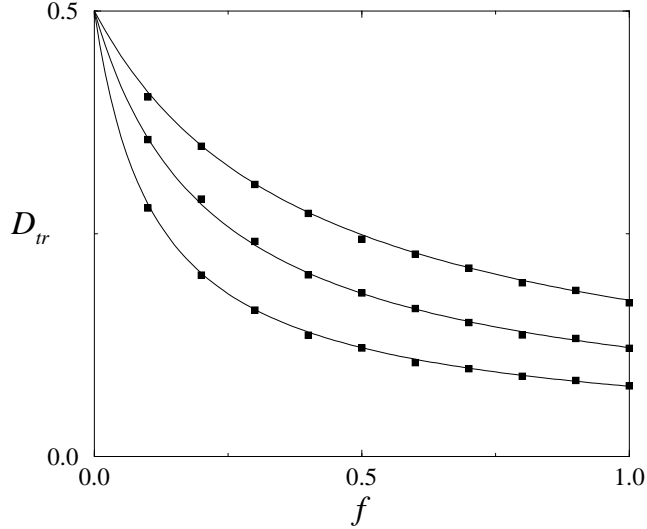}
\end{center}
\caption{\label{self} {\small The TP self-diffusion coefficient vs
the adsorption probability $f$. Notations and values of $g$ are
the same as in Figs.\ref{vitfig}.}}
\end{figure}

Consider finally the situation with $E = 0$, in which case the
terminal velocity vanishes and one expects conventional diffusive
motion such that
\begin{equation}
\label{diff}
\overline{X_{tr}^2(t)} =  2 D_{tr} t,
\end{equation}
where $D_{tr}$ is the diffusion coefficient. We can evaluate
$D_{tr}$ if we assume that the Einstein relation $ D_{tr}=\beta/
\zeta$ holds, which yields
\begin{equation}
\label{ddiffusion}
D_{tr} = \frac{\sigma^2 (1 - \rho_s)}{2 \tau} \left\{1 +\frac{\rho_s\tau^*}{\tau(f+g)}
\frac{2}{1 + \sqrt{1 +2 (1 + \tau^* (1 - \rho_s)/\tau) /(f+g)}} \right\}^{-1}.
\end{equation}
 Monte Carlo
simulations  evidently confirm our prediction for $D_{tr}$ (see
Fig.\ref{self}).

\vspace{0.3in}

{\bf RESULTS FOR TWO-DIMENSIONAL ADSORBED MONOLAYERS}

\vspace{0.2in}

We turn now to the case of a 2D substrate. In this case, the
general solution for the density profiles, in the frame of
reference moving with the TP, reads
\cite{benichou,physa,prl,benichouPRB}:
\begin{eqnarray}
k({\bf \lambda}) = k_{n_1,n_2} = \rho_s +
\alpha^{-1}\Big\{\sum_\nu A_\nu h({\bf e}_{\bf \nu})\nabla_{-\nu}F_{n_1,n_2}-
\rho_s(A_1-A_{-1})(\nabla_1-\nabla_{-1})F_{n_1,n_2}\Big\}
\label{2dsolh}
\end{eqnarray}
with
\begin{eqnarray}
F_{n_1,n_2}=
\left(\frac{A_{-1}}{A_1}\right)^{n_1/2}\int_0^{\infty}e^{-t}{\rm I}_{n_1}\left(2\alpha^{-1}\sqrt{A_1A_{-1}}t\right){\rm I}_{n_2}\left(2\alpha^{-1}A_2t\right){\rm d}t,
\label{repint}
\end{eqnarray}
where ${\rm I}_n(z)$ stands for the  modified Bessel function, the coefficients $A_\nu$  are
determined implicitly as the solution of the following system of
three non-linear matrix equations
\begin{equation}
\forall\nu=\{\pm1,2\},\;\;\;\;A_\nu=1+\frac{4\tau^*}{\tau}p_\nu\left\{1-\rho_s-\rho_s(A_1-A_{-1})\frac{\det\tilde{C}_\nu}{\det\tilde{C}}\right\},
\label{2dimplicite}
\end{equation}
\begin{equation}
\tilde{C}=
\begin{pmatrix}
A_1\nabla_{-1}F_{\bf e_1}-\alpha & A_{-1}\nabla_1F_{\bf e_1} & A_{2}\nabla_{-2}F_{\bf e_1}\\
A_1\nabla_{-1}F_{\bf e_{-1}} & A_{-1}\nabla_1F_{\bf e_{-1}}-\alpha & A_{2}\nabla_{-2}F_{\bf e_{-1}}\\
A_1\nabla_{-1}F_{\bf e_2} & A_{-1}\nabla_1F_{\bf e_2} &  A_{2}\nabla_{-2}F_{\bf e_2}-\alpha
\end{pmatrix}.
\end{equation}
the matrix  $\tilde{C_\nu}$ stands for the matrix obtained from  $\tilde{C}$
by replacing the $\nu$-th column  by the column-vector
$\tilde{F}$,
\begin{equation}
\tilde{F}=
\begin{pmatrix}
(\nabla_1-\nabla_{-1})F_{\bf e_1}\\
(\nabla_1-\nabla_{-1})F_{\bf e_{-1}}\\
(\nabla_1-\nabla_{-1})F_{\bf e_2}
\end{pmatrix},
\end{equation}
while $h({\bf e}_{\bf \nu})$ are expressed in terms of $A_\nu$ as
\begin{equation}
h({\bf e}_{\bf
\nu})=(1-\rho_s)+\frac{\tau}{4\tau^*p_\nu}(1-A_\nu).
\label{ppasnul}
\end{equation}
Lastly, we find that the TP velocity obeys
\begin{equation}
V_{tr}=\frac{\sigma}{\tau}(p_1-p_{-1})(1-
\rho_s)\Big\{1+\rho_s\frac{4\tau^*}{\tau}\frac{p_1\det\tilde{C}_1-p_{-1}\det\tilde{C}_{-1}}{\det\tilde{C}}\Big\}^{-1}.
\label{forcevitesse}
\end{equation}

Non-linear Eqs.(\ref{2dsolh}) are quite complex and their explicit
solution can not be obtained analytically. Typical density
profiles are depicted in Fig.5 and are characterized by a
condensed, traffic-jam-like region in front of the stationary
moving TP, and a depleted by particles region past the TP.

The asymptotical behavior of the density
profiles at large distances from
the TP is quite spectacular. In front of the TP,
the deviation $h_{n,0} = k_{n,0} - \rho_s$ always decays exponentially with the distance:
\begin{equation}
h_{n,0}\sim K_+\frac{\exp\Big(-n/ \lambda_+\Big)}{n^{1/2}}, \;\;\; \lambda_+ =
\ln^{-1}\Big(\frac{1}{A_{-1}}\left\{\frac{\alpha}{2}-A_2+\sqrt{\left(\frac{\alpha}{2}-A_2\right)^2-
A_1A_{-1}}\right\}\Big).
\end{equation}

On contrary, the behavior past the TP   depends qualitatively on
the physical situation. In the general case when  exchanges with
the vapor phase are allowed, the decay of the density profiles is
still exponential:
\begin{equation}
h_{-n,0}\sim K_-\frac{\exp\Big(-n/ \lambda_-\Big)}{n^{1/2}}, \;\;\; \lambda_- = - \ln^{-1}\Big(\frac{1}{A_{-1}}\left\{\frac{\alpha}{2}-A_2-\sqrt{\left(\frac{\alpha}{2}-A_2\right)^2-A_1A_{-1}}\right\}\Big)
\end{equation}

\begin{figure}[ht]
\begin{center}
\includegraphics*[scale=0.6,angle=0]{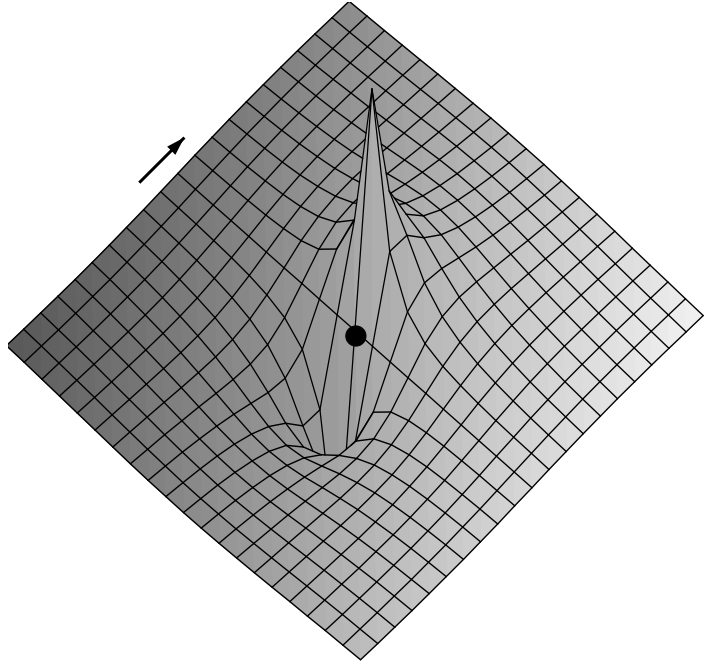}
\end{center}
\caption{\label{reseau} {\small Typical density profiles in 2D case. The arrow shows the direction of external field.}}
\end{figure}

In the conserved particles number limit, which can be realized for
the monolayers sandwiched in a narrow gap between two solid
surfaces, $\lambda_-$ becomes infinitely large and the deviation
of the particle density from the equilibrium value $\rho_s$
follows
\begin{equation}
h_{-n,0}=-\frac{K_-'}{n^{3/2}}\left(1+\frac{3}{8n}+{\mathcal
O}\Big(\frac{1}{n^2}\Big)\right). \label{algebrique}
\end{equation}
Remarkably enough, in this case the correlations between the TP
position and the particles distribution vanish  {\it
algebraically} slow with the distance! This implies
 that mixing of the monolayer is not efficient enough to prevent the
appearance of the quasi-long-range order and the medium
"remembers"
 the passage of the TP
on a long time and space scale.

Now, in the limit $\beta E \ll 1$ we find a Stokes-type formula of
the form $V_{tr}\sim E/ \zeta $, where
\begin{equation}
\label{rrr}
\zeta=\frac{4\tau}{\beta\sigma^2(1-\rho_s)}\left\{1
+\frac{\tau^*}{\tau}\frac{\rho_s}{\Big(f + g + 1 + \tau^* (1 - \rho_s)/\tau\Big)\Big({\cal L}(x)-x\Big)}\right\},
\end{equation}
with
\begin{equation}
x = \frac{1}{2} \frac{1 + \tau^* (1 - \rho_s)/\tau}{f + g + 1 + \tau^* (1 - \rho_s)/\tau} \;\;\; \text{and} \;\;\;
{\cal L}(x)\equiv\left\{\int_0^\infty e^{-t}\Big(({\rm I}_0(xt)-{\rm I}_2(xt)){\rm I}_0(xt){\rm d}t\right\}^{-1}.
\end{equation}
Note  that we again are able to single out two physically meaningful
contributions to the friction coefficient $\zeta$.  Namely, the
first term on the rhs of Eq.(\ref{rrr}) is just the
mean-field-type result corresponding to a perfectly stirred
monolayer, in which correlations between the TP and the monolayer
particles are discarded. The second term, similarly to the 1D
case, mirrors the cooperative behavior emerging in the monolayer
and is associated with the backflow effects.  In contrast to the
1D case, however, the contribution stemming out of the cooperative
effects remains finite in the conserved particles limit.

Lastly, we estimate the TP diffusion coefficient $D_{tr} =
\beta^{-1}\zeta^{-1}$ as
\begin{equation}
D_{tr}=\frac{\sigma^2}{4\tau}(1-\rho_s)\left\{1-\frac{2\rho_s\tau^*}{\tau}\frac{1}{ 4
(f + g + 1 + \tau^* (1 - \rho_s)/\tau) {\cal L}(x)-1+(3\rho_s-1)\tau^*/\tau}\right\}.
\label{2dautodiffgen}
\end{equation}
Note that setting $f$ and $g$ equal to zero, while assuming that
$f/g=\rho_s/(1-\rho_s)$, we recover from the last equation the
classical result due to Nakazato and Kitahara (see \cite{kehr}),
which is exact in the limits $\rho_s \ll 1$ and $\rho_s \sim 1$,
and serves as a very good approximation for the self-diffusion
coefficient in hard-core lattice gases of arbitrary density
\cite{kehr}.

\vspace{0.3in}

{\bf CONCLUSIONS}

\vspace{0.2in}

To conclude,  we have studied analytically the
 intrinsic frictional properties
 of adsorbed monolayers,
composed of mobile hard-core particles undergoing continuous
exchanges with the vapor. We have derived a system of coupled
equations describing evolution of the density profiles in the
adsorbed monolayer, as seen from the  moving tracer, and its
velocity $V_{tr}$.  We have shown that  the density profile around
the TP is strongly inhomogeneous: the local density of the
adsorbed
 particles in front of the TP is higher than the
average and approaches the average value as an exponential
 function of the distance from the TP.
On the other hand, past the TP the local density is always lower
than the average, and depending on whether the number of particles
is explicitly conserved or not, the local density past the TP
 may tend to the average value either as an exponential or even as an
 $\it algebraic$ function of the distance. The latter reveals
especially strong memory effects and strong correlations between
the particle distribution in the environment and the TP position.
Next, we have derived a general force-velocity relation, which
defines the TP terminal velocity  for arbitrary applied fields and
arbitrary values of other system parameters. We have demonstrated
next that in the limit of a vanishingly small external bias this
relation attains a simple, but physically meaningful form of the
Stokes formula, which signifies that in this limit the frictional
force exerted on the TP  by the monolayer particles is viscous.
Corresponding friction coefficient has been also explicitly
determined. In addition, we estimated the self-diffusion
coefficient of the tracer in the absence of the field.

\vspace{0.3in}

{\bf ACKNOWLEDGMENTS}

G.O. thanks the AvH Foundation for the financial support via the Bessel Research Award.

\vspace{0.2in}

\vspace{0.2in}

\renewcommand{\baselinestretch}{1.0}


\begin{thebibliography}{99}


\bibitem{spreading1} S.F.Burlatsky, G.Oshanin, A.M.Cazabat, and M.Moreau, Phys. Rev. Lett. {\bf 76}, 86 (1996);
Phys. Rev. E {\bf 54}, 3892 (1996)
%
\bibitem{aussere} D.Ausserr{\'e}, F.Brochard-Wyart and P.G.de Gennes, C. R. Acad. Sci. Paris {\bf 320}, 131 (1995)
%
\bibitem{dewetting} G.Oshanin, J.De Coninck, A.M.Cazabat, and M.Moreau,
Phys. Rev. E {\bf 58}, R20 (1998);  J. Mol. Liquids {\bf 76}, 195 (1998).
%
\bibitem{islands} see, e.g. H.Jeong, B.Kahng and D.E.Wolf, Physica A {\bf 245}, 355
(1997); J.G.Amar and F.Family, Phys. Rev. Lett. {\bf 74}, 2066 (1995) and references
therein
%
\bibitem{surf} M.-C.Desjonqu{\'e}res and D.Spanjaard, {\em Concepts in Surface Physics},
(Springer Verlag, Berlin, 1996)
%
\bibitem{wahn} R.Ferrando, R.Spadacini and G.E.Tommei, Phys. Rev. E {\bf 48}, 2437
(1993) and references therein
%
\bibitem{kehr} K.W.Kehr and K.Binder, in:
 {\em Application of the Monte Carlo Method in
Statistical Physics}, ed. K.Binder,
(Springer-Verlag, Berlin, 1987) and references therein.
%
\bibitem{nouspercol} O.B{\'e}nichou, J.Klafter, M.Moreau, and G.Oshanin,
Phys. Rev. E {\bf 62}, 3327 (2000)
%
\bibitem{lyklema} J.Lyklema, {\it Fundamentals of Interface and Colloid
Science}, Volume II (Solid-Liquid Interfaces), (Academic Press Ltd., Harcourt Brace, Publ., 1995)
%
\bibitem{benichou} O.B{\'e}nichou, A.M.Cazabat, A.Lemarchand,
 M.Moreau, and G.Oshanin, J. Stat. Phys. {\bf 97}, 351 (1999)
%
\bibitem{physa} O.B{\'e}nichou, A.M.Cazabat,
 M.Moreau, and G.Oshanin, Physica A {\bf 272}, 56 (1999)
%
\bibitem{prl} O.B{\'e}nichou, A.M.Cazabat, J.De Coninck,
M.Moreau, and G.Oshanin,  Phys. Rev. Lett. {\bf 84}, 511 (2000)
%
\bibitem{benichouPRB} O.B{\'e}nichou, A.M.Cazabat, J.De Coninck,
M.Moreau,  and G.Oshanin,  Phys. Rev. B {\bf 63}, 235413 (2001)
%
\bibitem{10} R.F.Steiner, J. Chem. Phys {\bf 22}, 1458 (1954)
%
\bibitem{bur2} S.F.Burlatsky,  G.Oshanin, M.Moreau, and W.P.Reinhardt, Phys. Rev. E {\bf 54},  3165 (1996)


\end{thebibliography}
\end{document}